\documentstyle[11pt]{article}
\newcommand{\ul}{\underline}
\author{R.M.Weiner
\\Physics Department, University 
of Marburg, Marburg, Germany
\thanks{E.-mail address: 
weiner@mailer.uni-marburg.de} 
\\and
\\Laboratoire de Physique Th\'{e}orique et Hautes Energies
\\Universit\'{e} de Paris-Sud, Orsay, France}
\date{}\title{Bose-Einstein Correlations\\A Research Program
for the 21-st century
\thanks{ Talk given at the 7th International Workshop 
on Multiparticle Production ``Correlations and Fluctuations"
Nijmegen, The Netherlands, June 1996}}

\begin{document}
\maketitle
\begin{abstract}
It is argued that Bose-Einstein correlations
 constitute one of the most important and characteristic
effects of strong interactions. The progress made in
  our understanding of this phenomenon  is
 reviewed and a program for future research 
in this field
is formulated. 
\end{abstract}

\section{The state of the art}
\subsection{Introduction}
Although the study of Bose-Einstein correlations (BEC) is going on for more than 40 
years and many interesting theoretical developments took place
the  experimental {\em facts} which we know at present are
very
few (cf. subsection 1.5) and       
 it will take many decades until this situation will
significantly improve.   
This is necessary not only because BEC per se 
are of high scientific interest 
 (this will hopefully
emerge also from the following)
but primarily because  
in analogy to the fact that the Hanbury-Brown Twiss effect
 lead to a new chapter in optics, namely 
Quantum Optics, one could expect that BEC will lead to
a new chapter in strong interaction physics. 

What can be done to achieve this goal?
 The purpose of this talk is to try to answer
this question by formulating a program which could become
an entry ticket to this new field.
 
The talk will be divided into two parts: in the first and main
part
I will try to summarize the present status of the field.
Due to space limitations this summary has to be very sketchy. 
In the second part the open and most important problems will be
enumerated.

\subsection{Why are BEC interesting and important?}
Intensity interferometry and 
BEC in
particular 
constitute at present the only
 experimental method for the determination of sizes and
lifetimes of sources in particle and nuclear physics. The
measurement of these is essential for an understanding of the
dynamics of strong interactions which are responsible for 
the existence and properties of atomic nuclei. Moreover a
new state of matter, quark matter, 
in which the ultimate constituents move freely,
is within the reach of
present accelerators or those under construction. The
confirmation that we have really seen this ``promised" new
state is intimately linked with the determination of its
space-time properties. Furthermore certain consequences
of the standard model which could not be tested directly
should be seen in BEC and one of the most actual tests of
this model related to the much hunted Higgs particle is
influenced by this effect.
Last but not least 
besides this ``applicative" aspect
of BEC, this effect has important implications for the 
foundations of quantum mechanics.
 
Because of these facts
in the last years there has been a considerable surge 
of interest in hadron intereferometry and in particular in 
 BEC. At present there is
no
meeting on multiparticle production where 
numerous contributions to
this subject are not presented and at least one meeting
was dedicated almost entirely to this topic (cf. 
\cite{camp})
. Due to its excellent programing and
 organization the 
present meeting is a further milestone in this evolution.   

\subsection {Relation between BEC and
 and quantum field theory}
Loosely speaking Bose-Einstein correlations
 can be viewed as a consequence of the
symmetry properties of the wave function
with respect to permutation of two identical particles
with integer spin and are thus intrinsic
quantum phenomena.  At a higher level, these symmetry properties
of identical particles are expressed by the commutation relations
of the creation and annihilation operators of
particles in the second quantisation (quantum field
theory-QFT).
The QFT is the more general approach 
as it contains the possibility to deal with creation and 
annihilation of particles and certain correlation phenomena
like the correlation between particles and antiparticles can
be properly described only within this formalism.
Furthermore, at high energies, because of the large number of particles
produced, not all particles can be detected in a given
reaction and therefore one usually measures only inclusive 
cross sections. For these reactions the wave function
formalism
is impracticable.
Related to this is the fact that the second quantisation
provides through the density matrix a transparent link
between correlations and multiplicity distributions. This
last topic has been in the center of interest of
multiparticle
dynamics for the last 20 years (we refer among other things
to KNO scaling and intermittency).
Last but not least one of the most important properties of
systems made of identical bosons and which is responsible
for the phenomenon of {\em lasing} is quantum statistical
coherence. This feature is also not accessible to a
theoretical
treatment except in field theory. 

\subsection {HBT versus GGLP. Final state interactions.}

The method of photon intensity interferometry was 
invented in
the mid fifties by Hanbury-Brown and Twiss for the
measurement
 of stellar dimensions and is sometimes called the
HBT method. In 1959-1960 G.Goldhaber, 
S.Goldhaber,W.Lee and A.Pais discovered 
that identical charged pions produced in $\bar p-p$ annihilation
are correlated (the GGLP effect). Both the HBT and the GGLP 
effects are based on BEC. In HBT we deal with correlations between photons, i.e.
particles which practically do not interact, while in GGLP
we have hadrons which interact. 
This fact made some people 
wonder whether hadron
interferometry is possible at all. Before going into a
theoretical discussion of this question it is useful to recall
  some qualitative experimental facts which suggest
that the above doubts are unjustified. These facts are:
a) Positive correlations are seen between identical particles in all reactions like e-e, hadron-hadron, lepton-
hadron, hadron-nucleus and nucleus-nucleus at all energies
and are not seen between non-identical particles, except
for resonance effects 
(the issue of quantum statistical particle-antiparticle
correlations is discussed in section 1.8).
b) BEC increase with decreasing difference of the momenta of
the pair
as expected in any theoretical treatment of BEC. 
c) Radii extracted via BEC from heavy ion reactions increase
with mass number of the participating nuclei.
It is these few facts which were alluded to above as being
the basis of our confidence into the fact that in the GGLP
effect we really see BEC.

\subsection {BEC and the search for quark gluon plasma}
The experimental proof of the existence of quark-gluon plasma
as a new phase of matter is certainly one of the most 
challenging 
tasks of high energy physics. BEC play in this game a double
role: a) To prove that QGP is indeed a (new) {\em phase} 
one must prove that
its lifetime is significantly larger than the typical
hadronic time scale ($10^{-23}$ sec).
Hence a lifetime measurement is necessary. b) To prove that a 
phase transition took place one must prove that the
energy density achieved in the reaction exceeds the critical
energy density predicted by statistical QCD. This implies
the
measurement of the volume of the fireball.
Both a) and b) can be performed only
with the help of BEC.
Unfortunately, most of the experimental results on BEC with
relativistic
heavy ions are still called ``preliminary".

\subsection {BEC and the foundations of quantum mechanics}
There are some aspects of principles of quantum mechanics
involved in the study of BEC, which have been discovered
more recently and which are related to the very concept
of {\em identity} of particles. It is well known that the 
principle of identity of particles is part of
the fundamental postulates of quantum mechanics and states
essentially that elementary particles are indistinguishable.  
The question what means {\em identical} has not been
raised until recently,
because it had been considered that the answer to it was
obvious. This situation has changed 
when it was discovered
that there exists also a quantum statistical correlation
between particles and antiparticles 
\cite{prl}.

Another fundamental property of BEC emerges from
 Bose statistics:
for small differences in the momenta of the pair, identical
bosons are in general {\em bunched} while fermions are 
{\em antibunched}.
However in BEC for the particular case of squeezed states
(cf. section 1.9) also antibunching is possible.

\subsection {BEC, quantum statistics and the standard model}
\subparagraph{Coherence}
While for the determination of sizes and lifetimes via
intensity interferometry, both bosons and fermions can be
used, BEC 
have another potential field of applications of major
interest for particle physics,
  namely the determination of the
amount of coherence of sources. This constitutes also a 
test of 
the presence
of classical fields (any classical field is a coherent
state). 
To realise the importance of this topic it is enough to
mention that some of the most important developments in 
particle physics
of the last 25 years including the ``standard model", are based on spontaneously broken
symmetries which imply classical fields. 
However so far there is no direct experimental evidence 
for these fields. On the other hand it is well known from
quantum optics that BEC depend on the amount of coherence     
in a very characteristic way (a completely coherent source
like a laser above threshold leads to a constant correlation
i.e. no bunching) and therefore one hopes to obtain
information
about coherence from boson interferometry.
\subparagraph{The density matrix}
This dependence of BEC on coherence is a particular case
of the fact that in
quantum theory any probability or cross section of an
operator $O$ is an expectation value and thus depends
 on the state of the system which   
in general
is described by the density matrix $\rho$
\begin{displaymath} <O> = Tr(\rho O). 
\end{displaymath}
$\rho $ is in principle determined by the theory. 
For hadron multiproduction this theory is quantum
chromodynamics and for processes involving multiphoton
production this theory is
quantum electrodynamics. However in both cases the use of
the fundamental theory is impracticable because of the
complexity of the many body problem. That is why one has to 
use
 phenomenological approaches. 
 The experience gained in photonics
has been instrumental in the analogous problem of hadron
multiparticle production and amounts essentially to
postulating
the form of the density matrix in the coherent state
representation $|\alpha>$ .    
Let us consider a statistical distribution
${\cal P} \{ \alpha \}$
  and expand the density matrix
in $|\alpha>$. 
Given the fact that the coherent states form an
(over)complete set, this means that the resulting 
density matrix $\rho$
is quite general. Indeed we write then 
\begin{displaymath}
\rho = \int {\cal D} \alpha \ {\cal P}\{ \alpha \} \ |\alpha><\alpha|
\label{eq:what2}
\end{displaymath}
where the symbol ${\cal D}\alpha$ denotes an integration 
over the space 
of functions $\alpha$, and the statistical weight 
${\cal P}\{ \alpha \}$ is 
normalized to unity.

The simplest and most common form of density
matrix used is the Gaussian form in the $\cal P$
representation
 both because of its
mathematical convenience as well as because of the fact that
it corresponds to the physically important case when the
number of independent sources is large (central limit theorem).
Two important phenomenological consequences follow from the
Gaussian form of $\rho$.

(i) The maximum of the BEC function is quantitatively 
well defined, independent of the concrete form of the
field correlator and of the geometry; thus 
e.g. for the second order BEC function $maxC_{2} = 2$.

(ii) The density matrix can be expressed in terms of only
two
moments of ${\cal P}$
(cf.below).
Most experimental data so far, with the exception of
annihilation
in rest (cf. \cite{sqAW} and references quoted there), are
consistent with (i). As to (ii) the situation is less clear.
An approximate confirmation of (ii) has been obtained in
\cite{evid} based on the data of \cite{neum}.
While it appears that the gaussian ansatz is at least an
acceptable
approximation, given the importance
of
the form of the density matrix, more precise tests are very
important. Furthermore, the consequences of small deviations
from this form for the relationship between correlation
functions of different order and for the relationship
between moments of the 
multiplicity distributions of different order could be theoretically worked
out without difficulties. This would
 make the form of the density matrix more accessible
to
experimental tests.

\subparagraph{The current formalism}   
In particle physics 
rather than working with the fields it is often
convenient and sufficient to use classical currents $J$ and
this will be done in the following.
 One can show that this effectively amounts to substitute in the above
eqs. the symbol $\alpha$ characterizing the coherent field
by the symbol $J$ characterizing the classical current.

The current formalism has two important advantages:
1. The corresponding field theoretical Klein Gordon equation 
can be solved exactly. 
2. The space time picture of the process we are
interested in can be introduced immediately.
The current $J(x)$ can generally be written as the sum of 
a chaotic and a coherent component,
\begin{displaymath}
J(x) \ = \ J_{chaotic}(x)\ + \ J_{coherent}(x)
\label{eq::jsum}
\end{displaymath}
with 
\begin{eqnarray*}
J_{coherent}(x) & = & <J(x)>,\\
J_{chaotic}(x)  & = & J(x) - <J(x)>. 
\label{eq::jchao}
\end{eqnarray*}

By definition, $<J_{chaotic}(x)>=0$.
The case $<J(x)> \not= 0$ corresponds to {\em single particle
coherence}.
In the following we shall also deal with {\ two-particle 
coherence} 
(squeezed
states).

A Gaussian current distribution is completely determined by
specifying its first and second moments: the coherent component,
\begin{displaymath}
I(x) \ \equiv \ <J(x)>
\label{eq:a1}
\end{displaymath}
and the 2-current correlator,
\begin{eqnarray*}
D(x,x') & \equiv & <J(x) \ J(x')> - <J(x)> \ <J(x')>\nonumber\\
  & = & <J_{chaotic}(x) \ J_{chaotic}(x')>.
\label{eq:a2}
\end{eqnarray*}

Consider first the case of an infinitely extended source. The 
correlation of currents at two space-time points $x$ and $y$ 
is described by a primordial correlator,
\begin{displaymath}
<J(x) \ J(y)>_0\  =\  C(x-y).
\label{eq:jx}
\end{displaymath}
The correlator $C(x-y)$ reflects dynamical properties of
the particle
source, rather than its space-time geometry. 
Effects of the geometry of the source 
are taken into account by introducing
the space-time distributions of the chaotic and of the coherent component,
$f_{ch}(x)$ and $f_c(x)$, re\-spec\-ti\-ve\-ly. The expectation values of the 
currents, $I(x)$ and $D(x,x')$, take nonzero values only in space-time 
regions where $f_c$ and $f_{ch}$ are nonzero. Thus, one may write 
\begin{eqnarray*}
I(x) &=& f_c(x)\\
D(x,x') &=& f_{ch}(x) \  C(x-x') \ f_{ch}(x').
\label{eq:ixh}
\end{eqnarray*}

A microscopic
theory like lattice QCD may eventually provide us with the exact
form of $C$. 
Phenomenologically one proceeds however by postulating the
analytical {\em form} of the correlator and parametrising
this form. General theoretical considerations allow then to
determine the minimum number of parameters. 
The next step is to describe the space-time distribution of
the source $f(x)$. Here again an analytical form has to be 
postulated and the number of independent parameters
determined
from general considerations
\footnote{For
mathematical simplicity usually Gaussian forms for $f$ and
$C$
are chosen. Contrary to the case of the density
matrix,
where the Gaussian form has, because of the central limit
theorem, deep theoretical significance, this is not true for
the
correlator $C$ or the geometrical function $f$.}.

The primordial correlator $C$ contains in general two 
 length scales (correlation lengths) characterising the two
space directions, e.g. the longitudinal and transverse
direction
for a non-expanding source or the boost and transverse
direction
for an expanding source, and one time scale (correlation 
time).  
Next we come to the space-time distributions of the source
$f(x)$ for which we have to distinguish between the chaotic
and coherent part $f_{ch}$ and $f_{c}$ respectively. 
For each of these parts we have again to
postulate the analytical form and to determine the
minimum number of independent parameters. One can show that
here again at least three parameters are necessary both for
expanding and non-expanding sources. 
Last but not least the chaoticity 
\begin{displaymath}
p(k) \ =\ \frac{D(k,k)}{D(k,k)+|I(k)|^2}
\label{eq:fk}
\end{displaymath}
has to be determined. One
can show that within the above classical current formalism
this can be done by specifying the value of $p$ at $k=0$.
This means that all overall there are at least 10 independent
parameters which have to be determined phenomenologically
in order to characterize completely BEC \cite{intj}
\footnote{A semiclassical approximation of the current formalism 
is the Wigner
function formalism. It is useful when combined with full
fledged
hydrodynamics as it provides a link between correlations
and the 
equation of state \cite{Bolz}, a subject of high current
interest in the investigation of hadronic and quark matter.
This combination is apparently powerful enough to explain
a multitude of heavy ion physics data, including single and
double inclusive cross sections, particle yields etc.
Heuristically
it also has the advantage that it makes assumptions about
the forms of the primordial correlator $C$ and of the 
geometry $f$   
unnecessary as these follow from the solutions of the
equations
of hydrodynamics. 

On the other hand in the heavy ion physics literature
sometimes ``short cuts" of the Wigner formalism 
are used in which hydrodynamics is replaced by assumptions
about the form of the {\em source} function which are
 equivalent
to the assumptions made above about $C$ and $f$. However,
 not only is this last approach
 less economical than the current
formalism in the form presented above, (the number of
independent parameters for a chaotic source alone is ten;
these parameters are dependent on the average momentum of
the pair and this dependence is not predicted by the 
``theory"),
but it is also less general as it is based on a 
semiclassical approximation, valid only for small values
of the difference of momenta $q$.}.

\subsection{Particle-Antiparticle correlations}

A surprising consequence of the classical current formalism, which
however holds also for quantum currents \cite{qft} is the
existence of particle-antiparticle correlations and the fact
that BEC for neutral pions are different from BEC for
charged ones. The experimental detection of these new
effects which are quite small but nevertheless very
important
from a principle point of view demands among other things 
 much higher statistics
at small momentum differences than that achieved so far. For
details of the derivation of these effects we refer the
reader to \cite{prl},\cite{intj}. Qualitatively these effects can be 
understood if one realises that in the 
s-channel the particle-antiparticle interaction is different
from the charged-charged interaction, because of the
annihilation channel.

\subsection{Squeezed states}
At this point one should also mention that besides
ordinary
coherent states used as the basis of the representation,
squeezed coherent states have been introduced, which 
are of major interest both from a theoretical point of view 
as well as because of their application potential.
BEC are one of the most sensitive tools for 
the detection of these squeezed states. 
So far they have been seen only in optics mainly because 
in particle
physics there was missing a definite prescription how to 
produce them. However very recently it has been pointed out
that they may be produced
 in sudden transitions like annihilation
processes or the explosive hadronisation of
a quark gluon plasma \cite{sqAW}.

\subsection {BEC and multiplicity distributions.}
One of the most characteristic properties of strong 
interactions is many 
particle production. This in itself explains from the
beginning the need and usefulness of statistical methods.
The pertinant physical observable is
first of all the multiplicity distribution $P(n)$ which is
given by the diagonal matrix elements of the density matrix
in the number representation

\begin{displaymath}
P(n) = <n|\rho|n>.
\end{displaymath}

 One classifies
correlations in strong interactions into long range (LRC) and
short range (SRC). If one restricts oneself to identical bosons
the bulk of SRC is due to BEC. However LRC also influence
BEC ; this was seen in hadronic reactions \cite{LRC}
but in $e^{+}-e^{-}$
reactions
and heavy ion reactions the situation is less clear.  
 
As long as one considers $P(n)$ in the entire phase space
they
are dominated by LRC, while if one restricts the phase space
to
small windows (e.g. in rapidity) SRC play the main part. The
fact that SRC for identical particles are essentially due
to BEC has led to the proposal \cite{Fowler}  
 to combine the investigation of multiplicity
distributions with that of BEC and to look for coherence
effects simultaneously in $P(n)$ and BEC. 
 
In 1986 Bialas and
Peschanski \cite{BP} suggested
that the moments of $P(n)$ in narrow windows might scale 
and show an {\em 
intermittency} pattern. Although this interpretation of the
data available at the end of the eighties had been disputed
quite soon when it was pointed out that conventional SRC
analogous to the quantum statistical ones might be
responsible
for these observations \cite{Carr}, the link with BEC was not accepted 
definitively 
until it was proven experimentally that the so called scaling
effects were strongly enhanced if only identical particles   
were analysed. 
Subsequently
Bialas \cite{Bi} suggested that the source itself might be
fractal; this would
 explain the above new experimental observations without
spoiling the ``intermittency" interpretation.

As demonstrated however in \cite{39}
the apparent scaling behaviour can be explained by
conventional BEC with a source of {\em fixed} size.
 Nevertheless a more definitive clarification of this 
issue awaits better resolution at
small $q$ where an end of the present ``scaling" is
predicted by the conventional BEC theory.

Another aspect of the relationship between BEC and
multiplicity distributions is the fact that
 $P(n)$ depends not only on the width of the rapidity region
but also on the position on the
rapidity
axis y: in the center $P(n)$ is broader
 than in the fragmentation region. Taking over the quantum
statistical
language as explained above this suggests that $P(n)$ is
more chaotic in the center \cite{FFW},\cite{W}. If this is true it should be
seen in BEC and indeed there are some experimental hints in
this direction \cite{B}. However much more experimental
 work and in particular a drastic improvement of statistics,
as well as the theoretical investigation of alternative 
explanations   
of the presumed observations are necessary before more
stringent conclusions can be drawn.

\subsection { Experimental problems} 
From the facts mentioned above the reader may realize that
there has been considerable progress in our understanding
of the phenomenon
 of BEC. Unfortunately these theoretical developments 
have not
been matched in full with comparable experimental progress
although the number of experimental papers on this subject
is quite impressive. Some of the most important reasons for
this defficiency are of technical nature like 
 insufficient statistics, lack of
particle and track by track
identification, and limitations of the phase space accessible
to detectors. As a matter
of fact many of the experiments on BEC performed so
far are not {\em dedicated\/} experiments but rather
byproducts of other experiments.
  
Another difficulty in the experimental investigation of BEC
is related to the problem of normalisation of the
correlation
functions. 
 Correlation functions of order $n$ are by definition 
ratios of
 multiple inclusive cross sections of order $n$ and 
the $n$ fold product of 
single inclusive
cross sections. Because of the phase space limitations of
most detectors used at present, the single inclusive
cross sections cannot be measured directly. To circumvent
this difficulty ``substitutes" for the denominator made of the
products of single inclusive cross sections are used. As any
substitute they are not ideal and introduce biases in the
experimental results. This is particularly evident in recent
measurements of BEC in $e^+-e^-$ reactions at the CERN LEP
accelerator where the values for the incoherence factor 
$\lambda$
obtained by different methods of normalisation at different
detectors differ by factors up to 2. Because of this 
situation serious doubts have been expressed about
the usefulness of these measurements \cite{Hay}.

\section{The program}

Based on the considerations of the previous section
I would like to propose the following research 
program in 
the field of BEC. Given the discrepancy between
theoretical and experimental progress, this program is
mostly of experimental character and will be labeled by (E). 
Wherever appropiate, 
theoretical open problems will also be mentioned and labeled
by (T).  
\begin{enumerate}
\item Determination of the form of the density matrix in the
${\cal P}$ representation (possible deviations from the
gaussian form), mostly from higher order BEC (E+T).  

\item Determination of of all independent 
parameters, in particular separation of correlation lengths
(times) from geometrical scales and determination of
chaoticity (E). Comparison with results to be obtained in
the mean time by lattice QCD and other theoretical
developments (T).

\item Determination of the form of the correlator and of the
form of the space
time distribution (E); comparison with future results from
lattice QCD and other theoretical approaches (T).

\item Comparison of BEC in e-e, hadron-hadron and heavy ion
reactions, with particular emphasis on 1, 2, and 3 (E+T).

\item Measurement of particle-antiparticle correlations (E).

\item Simultaneous determination of BEC and multiplicity
distributions in the same phase space region (E).

\item Search for squeezed states, both in BEC 
(through overbunching and antibunching) and in
multiplicity distributions (through oscillations in $P(n)$) (E).

\item Normalisation of BEC using the separately determined
single inclusive cross sections in the entire phase space (E).

\item Track by track detection and improved identification of
particles (E).

\item Improvement of statistics especially at small $q$
by at least 2 orders of magnitude (E). 
\end{enumerate}
The entire program could be summarised in
two words: {\em Quantum Hadronics}. 
This name
on the one hand reflects the 
analogy with quantum optics 
and on the
other
the fact that hadrons are not photons.
The implementation of this program may not take 100 years, but
it is certain that it will not be realized
 this century; therefore the title of this talk. Once the
above points and in particular 1-3 will be clarified, one
might be able to proceed to a reconstruction of an effective
density matrix for multiparticle production in strong interactions along the lines of
\cite{Botke},\cite{Fowler}. 
This could be then the birth
certificate
of the new chapter of physics alluded to in the introduction.

{\bf Acknowledgements}

I am indebted to the organizers of this workshop and in
particular to Wolfram and Susanne Kittel for their
warm and kind hospitality.

\end{document}